\documentclass[%
 reprint,
 amsmath,amssymb,
 aps,
]{revtex4-2}

\usepackage{comment}
\usepackage{color}
\usepackage{graphicx}
\usepackage{dcolumn}
\usepackage{bm}
\usepackage{hyperref}

\newcommand{\norm}[1]{\left\lVert#1\right\rVert}

\newcommand{\tr}{\text{tr}}

\newcommand{\modelnamefull}{\text{Lagrangian Attention Tensor Network }}
\newcommand{\modelabbr}{\text{LATN}}

\newcommand{\taueta}{\tau_\eta}

\begin{document}


\title{Lagrangian Attention Tensor Networks for Velocity Gradient Statistical Modeling}

\author{Criston Hyett $^{(1,2,*)}$}
\author{Yifeng Tian $^{(2)}$}%
\author{Michael Woodward $^{(2)}$}%
\author{Mikhail Stepanov $^{(1)}$}%
\author{Chris Fryer $^{(3)}$}%
\author{Daniel Livescu $^{(2)}$}%
\author{Michael Chertkov $^{(1)}$}%

\affiliation{
$^{(1)}$ University of Arizona, Applied Mathematics \& Mathematics \\
$^{(2)}$ Los Alamos National Laboratory, CCS Division \\
$^{(3)}$ Los Alamos National Laboratory, CNLS\\
$^{(*)}$ Corresponding Author: criston@hyett-research.org}

\date{\today}

\begin{abstract}
Direct numerical simulation of turbulence at realistic Reynolds numbers is still beyond current computational capability, necessitating models that reduce the number of resolved spatial scales.
Improving these models requires better models of the statistical dynamics of the Velocity Gradient Tensor (VGT) at the Kolmogorov scale.
Physics-Informed Machine Learning (PIML) has found considerable success in exploiting large datasets generated from direct numerical simulation (DNS) of Navier-Stokes to improve models for the evolution of the VGT.
Using the intimate connection between VGT evolution and fluid deformation, we augment a Tensor Basis Neural Network (TBNN) closure with learned Lagrangian memory kernels, and name the result Lagrangian Attention Tensor Networks (LATN).
Trained on single-time (tangent) data, Lagrangian memory is strongly predictive: it lowers the error of the pressure Hessian by 43\% and markedly improves its eigenvector alignment, with the most useful history spanning about half a Kolmogorov time.
These \textit{a-priori} gains, however, do not translate into improved \textit{a-posteriori} prediction.
We therefore inject the memory features into the first layer of a frozen, pre-trained TBNN and train \textit{only the memory} on model rollouts, against a distributional objective.
We integrate the model for nine Kolmogorov times from DNS initial conditions and evaluate it with single- and multi-time statistics of the DNS. The resulting LATN retains the trunk's single-time accuracy while improving the long-time statistics: the topology, magnitude and tail distributions of the VGT, and the alignment of vorticity with the strain rate.
\end{abstract}


\maketitle


\graphicspath{{figs_rev/}{figs/}}

\section{Introduction} \label{sec:intro}

The study of the Velocity Gradient Tensor (VGT), a classic and pivotal object in understanding turbulence, began with early closure attempts \cite{vieillefosse1984,cantwell1992exact}, leading to significant insights through simulations \cite{ashurst1987alignment} and experiments \cite{tsinober1992}. A key observation from this work was the preferential alignment of vorticity with the intermediate eigenvector of the strain rate, which the restricted Euler (RE) model, obtained by neglecting the anisotropic part of the pressure Hessian, reproduces qualitatively. However, it soon became evident that the RE closure, while stimulating further research, was fundamentally flawed, as it led to an unphysical finite-time singularity. To address this issue, the tetrad model \cite{chertkov1999lagrangian} took a Lagrangian approach, resisting the singularity by using a tetrad of tracer particles to track deformation along a Lagrangian trajectory.
Because the tetrad defines a coarse-grained velocity gradient, the model also built a bridge to the classic study of energy cascades initiated by Kolmogorov in 1941 \cite{kolmogorov1991local}, and significantly advanced since \cite{frisch1995}.

However, the Lagrangian closure assumptions in \cite{chertkov1999lagrangian} were difficult to verify directly in simulations, prompting the development of phenomenologies that avoided resolving spatial evolution of shapes. The new approaches suggested closures primarily in terms of local Lagrangian objects -- specifically incorporating the so-called Recent Fluid Deformation (RFD) approximation \cite{chevillard2006lagrangian}. See \cite{meneveau2011lagrangian} for an overview of the RFD phenomenology and its connection to the earlier stochastic diffusion \cite{girimaji1990diffusion}, linear damping \cite{martin1998dynamics}, and linear diffusion  \cite{jeong2003velocity} models. Development of the deformation tensor-based closure modeling of the pressure Hessian and viscous terms was also inspired by advances in Lagrangian simulations, particularly the challenging task of converting Eulerian Direct Numerical Simulation (DNS) data into Lagrangian data, as discussed in \cite{wan2016johns}. These datasets have enabled more in-depth study and closure modeling of the VGT, which is crucial for addressing unresolved problems in turbulence at high Reynolds numbers (Re), where DNS remains computationally unattainable. Such developments are important for improved sub-grid models in Large Eddy Simulations (LES) \cite{silvis2017physical}. This combination of data availability at comparatively low Re, the need for closure models at high Re, and the conjectured universality of VGT statistics has inspired renewed interest in Lagrangian VGT modeling \cite{das2020characterization,das2022effect,johnson2024multiscale}.

Throughout this development, the analysis of DNS datasets has played a crucial role in motivating and testing model hypotheses. This history of data-driven synthesis naturally aligns with advances in Machine Learning (ML). Specifically, Neural Network (NN)-based methods have emerged as powerful tools for analyzing Eulerian turbulence, as demonstrated by studies such as \cite{ling2016reynolds,wang2017physics,king2018deep,maulik2019subgrid,duraisamy2019turbulence,portwood2019turbulence,mohan2023embedding}. Approaching from the Lagrangian perspective, recent studies show the capability of physics-informed machine learning to model Lagrangian turbulence \cite{tian2021physics,tian2023lagrangian,woodward2023physics}.

Some NN-based approaches to turbulence modeling, such as those in \cite{wang2017physics,maulik2019subgrid,duraisamy2019turbulence}, have been predominantly data-driven, with physical symmetries, constraints, and interpretations incorporated retrospectively. In contrast, other methods, including those in \cite{ling2016reynolds,portwood2019turbulence,mohan2023embedding,tian2021physics,tian2023lagrangian,woodward2023physics}, were physics-informed in their structure. Both types of approaches align with the broader framework of Physics-Informed Machine Learning (PIML), which extends beyond turbulence applications --- see \cite{chertkov2024mixing} for a broader discussion of PIML and its importance in leveraging ML and AI to reveal the statistical nature of turbulence, complementing the classic pre-AI perspective \cite{jimenez2000large}. 

Recent models take the statistics of Lagrangian turbulence as their primary target \cite{carbone2024tailor,das2023data}: they reproduce these statistics by construction and use the remaining freedom in the formulation to incorporate physical properties, with impressive agreement with DNS. The data-driven closures closest to ours, the Tensor Basis Neural Networks (TBNN) applied to the VGT \cite{tian2021physics}, belong to a different lineage. They are trained on the single-time ($L_2$) error of the closure, i.e., for short-time accuracy of the dynamics, and they are \textit{Markovian}: the unclosed pressure Hessian and viscous terms are modeled as functions of the instantaneous VGT alone. The phenomenological models that preceded them (the tetrad, Recent Fluid Deformation and Recent Deformation of Gaussian Fields models) are built instead on -- approximations of -- the fluid element's \textit{deformation history}.

In this work we extend the TBNN lineage toward statistical closure. We incorporate the Lagrangian history into a PIML architecture, letting the data determine the importance and utility of the memory, and use it to begin closing the gap between closures trained for short-time accuracy and closures built to reproduce long-time statistics.

Toward this end, we augment the TBNN with Lagrangian memory kernels, a parameterized generalization of the linearized deformation tensor, and coin the result the \modelnamefull (\modelabbr). Memory is strongly predictive of the closure at single times, resolving much of the ambiguity of a closure conditioned on the current VGT alone. Trained on single-time error, however, the memory-augmented network degrades the integrated statistics; trained instead on model rollouts against a distributional objective, the memory improves them. Optimizing short-time fidelity and long-time statistical accuracy simultaneously will require dedicated training frameworks.

The rest of the manuscript is organized as follows: Section \ref{sec:prev_work} presents background along with the two motivating approaches, Section \ref{sec:methods} presents the \modelabbr\ model, its connection to previous Lagrangian deformation approaches and its two-stage training, Section \ref{sec:results} the \textit{a-priori} and \textit{a-posteriori} results, and Section \ref{sec:conclusion} the conclusions and open problems.

\section{Previous Work} \label{sec:prev_work}
\subsection{Governing Equations for VGT}\label{sec:governing}
The incompressible Navier-Stokes equations in three dimensions, ${\bm x}=(x_i|i=1,2,3)$ define the evolution in time $t$ of a velocity field ${\bm u}({\bm x},t)=(u_i({\bm x},t)|i=1,2,3)$
    \begin{gather} \label{eq:NS}
      \frac{\partial u_i}{\partial t} + u_j \frac{\partial u_i}{\partial x_j} = -\frac{\partial p}{\partial x_i} + \nu \frac{\partial^2 u_i}{\partial x_k \partial x_k} +f_i,
    \end{gather}
where we use Einstein notations (summation over repeating indexes), the velocity is incompressible, $\partial_{x_i} u_i=0$ (and the density of the fluid is set to $1$), and ${\bm f}({\bm x},t)=(f_i({\bm x},t)|i=1,2,3)$ is a large scale driving force of turbulence.
Applying spatial derivatives to Eq.~(\ref{eq:NS}), using the definition of material derivative and setting aside the (large-scale) driving force, whose gradient is negligible at the Kolmogorov scale and whose effect on the VGT is represented in the model by the stochastic term of Eq.~(\ref{eq:parameterized_sde}), we obtain an ODE for the VGT ${\bm A}({\bm x},t)\doteq(A_{ij}({\bm x},t)=\frac{\partial u_i}{\partial x_j}|i,j=1,2,3)$:
\begin{equation}
  \frac{dA_{ij}}{dt} = \frac{\partial A_{ij}}{\partial t} + u_k \frac{\partial A_{ij}}{\partial x_k} = - A_{ik}A_{kj} - \frac{\partial^2 p}{\partial x_i \partial x_j} + \nu \frac{\partial^2 A_{ij}}{\partial x_k \partial x_k}, \label{eq:vgt_ns}
\end{equation}
where transition from $\partial_t$ to $d/dt=\partial_t+{\bm u}\nabla_{\bm x}$ emphasizes our interest in evaluating ${\bm A}$ along Lagrangian trajectories -- that is in Lagrangian frame.
Taking the trace of Eq.~(\ref{eq:vgt_ns}) and using the incompressibility condition, we re-write the evolution equation for the VGT in the Lagrangian frame as
\begin{align} 
  \frac{dA_{ij}}{dt} = &\underbrace{\left(\frac{1}{3}A_{kl}A_{lk}\delta_{ij}-A_{ik}A_{kj}\right)}_{E_{ij}  \text{ (RE) }} \nonumber \\
  &+ \underbrace{\left(\frac{1}{3}\frac{\partial p}{\partial x_k \partial x_k}\delta_{ij}  -\frac{\partial^2 p}{\partial x_i \partial x_j}\right)}_{H_{ij} \text{ (PH) }} + \underbrace{\nu \frac{\partial^2 A_{ij}}{\partial x_k \partial x_k}}_{V_{ij} \text{ (VL) }}. \label{eq:vgt_evolution}
\end{align}
This equation has three traceless contributions:
the purely local (only in terms of ${\bm A}$ and not including spatial derivatives of ${\bm A}$) ``Restricted Euler'' (RE) term, $E_{ij}$, the deviatoric, nonlocal Pressure Hessian (PH) term, $H_{ij}$, and the (also nonlocal) Viscous Laplacian (VL) term, $V_{ij}$.

\subsection{Local Closures for the Evolution Equations}

While the viscous term is dynamically important, historically effort was focused on modeling the deviatoric pressure Hessian as it directly resists the RE-induced finite time singularity \cite{martin1998dynamics,tian2021physics}. Therefore, in this section, we review the literature from the perspective of closing the Hessian term, and we will consider the viscous term separately (in Section \ref{sec:VT}).

\subsubsection{Phenomenological Lagrangian Deformation Models} \label{sec:phenomenological_models}
In the first attempt to locally close the VGT equations, the so-called Restricted Euler dynamics were studied:
\begin{equation}
    \frac{dA_{ij}}{dt} \approx E_{ij}
\end{equation}
and an asymptotic analysis found this reduction induced a finite-time singularity \cite{vieillefosse1984internal}. This was subsequently confirmed with specification of exact solutions, invariant dynamics and topological evolution under the RE dynamics \cite{cantwell1992exact,cantwell1993behavior,li2007material}.

The universal behavior of fluid elements induced by the purely local RE approximation was one motivation for the introduction of the Tetrad Model \cite{chertkov1999lagrangian} that used four points to track and accumulate the history of the strain on a fluid element. This history was then utilized to close the nonlocal terms in Eq.~(\ref{eq:vgt_ns}), directly opposing the extreme flattening of the element and the acceleration of the phase plane dynamics along the Vieillefosse tail. 
The RFD model was (in part) motivated by the tetrad model under a short-history approximation \cite{chevillard2006lagrangian}.
The RFD model - and subsequent models in this family - utilized the structure of the pressure Hessian and viscous Laplacian, preferring to specialize models for each nonlocal term rather than attempt to close the combination of the two as done in the tetrad model. Finally, along this line of postulation and enriching the upstream condition via fitting coefficients of a truncated tensor expansion - introduced originally by Pope \cite{pope1975more} - the Recent Deformation of Gaussian Fields (RDGF) model was created \cite{johnson2016closure} (the tensor expansion in question is treated in detail in Section \ref{sec:tbnn}).

Pervasive in this lineage is the idea that the deformation history of a fluid element can inform and improve Markovian models. To formalize this concept, we consider, following \cite{chevillard2006lagrangian}, the Lagrangian path map, ${\bm X}\in\mathbb{R}^3 \mapsto {\bm x}(t)\in\mathbb{R}^3$, which provides the Eulerian position ${\bm x}$ at time $t$ of a fluid particle that was initially at position ${\bm X}$ at time $t_0$. Due to incompressibility, this map is invertible, and its Jacobian, the deformation (gradient) tensor ${\bm D} = \left(D_{ij} = \partial x_i/\partial X_j\bigg| i,j=1,2,3\right)$, satisfies $\text{det}({\bm D}) = 1$ at all times. The time evolution of the deformation tensor is governed by the differential equation:
\begin{equation}\label{eq:deformation_tensor_ode}
    \frac{d D_{ij}}{dt} = A_{ik} D_{kj}, \quad \text{with } D_{ij}(0) = \delta_{ij},
\end{equation}
where ${\bm A}$ represents the previously introduced VGT. The deformation tensor accumulates the Lagrangian history of the fluid element.

\subsubsection{Data-Driven Lagrangian Closures} \label{sec:tbnn}

We now return to the challenge of closing Eq.~(\ref{eq:vgt_evolution}) through the approximation of the deviatoric Pressure Hessian (PH), $H_{ij}$, using data-driven methods. For incompressible turbulence, a formal nonlocal expression for the deviatoric pressure Hessian can be derived as an integral over the spatial domain \cite{ohkitani1995nonlocal}:
\begin{equation} \label{eq:nonlocal_solution}
    H_{ij}({\bm x}) = \mathrm{P.V.}\!\iiint \frac{\delta_{ij} - 3\bar{r}_i \bar{r}_j}{2\pi |{\bm r}|^3} Q({\bm x} + {\bm r}) \, d {\bm r},
\end{equation}
where $\mathrm{P.V.}$ denotes the principal value of the singular integral, $Q = \frac{1}{2} A_{ij}A_{ji}$; ${\bm x}, \bm{r} \in \mathbb{R}^3$ and $\bar{\bm r} = {\bm r}/|{\bm r}|$. This representation highlights the nonlocal nature of the deviatoric pressure Hessian and its dependence on the surrounding VGT field.

It is well-established \cite{pope1975more,spencer1960further} that any tensor-valued function \( {\bm F}: \mathbb{R}^{3\times 3} \to \mathbb{R}^{3 \times 3} \) of the VGT can be expressed as a Taylor series expansion over the strain rate and rotation rate tensors:
\begin{equation*}
  {\bm F}({\bm A}) = \sum_{m,n = 0}^\infty \alpha_{mn} {\bm S}^m {\bm W}^n,
\end{equation*}
where
\begin{equation*}
  S_{ij} = \frac{1}{2}(A_{ij} + A_{ji}), \qquad W_{ij} = \frac{1}{2}(A_{ij} - A_{ji}),
\end{equation*}
are components of the symmetric strain rate tensor and the antisymmetric rotation rate tensor, respectively. This infinite series can be cast into a finite sum using the Cayley--Hamilton theorem applied to \( {\bm A} \).  Further, if the tensor function \( {\bm F}({\bm A}) \) is specified to be traceless, symmetric, and rotationally invariant, it can be represented using a tensor basis expansion \cite{zheng1993representations}:
\begin{equation} \label{eq:tensor_expansion}
  {\bm F}({\bm A}) = \sum_{n=1}^{10} g^{(n)}(\lambda_1, \dots, \lambda_5) {\bm T}^{(n)},
\end{equation}
where \( g^{(n)}: \mathbb{R}^5 \to \mathbb{R} \) are scalar functions of the five scalar invariants of the VGT:
\begin{align} \label{eq:vgt_invariants}
  \lambda_1 = \mathrm{tr}({\bm S}^2), \quad \lambda_2 = \mathrm{tr}({\bm W}^2), \quad \lambda_3 = \mathrm{tr}({\bm S}^3),\nonumber\\
  \lambda_4 = \mathrm{tr}({\bm W}^2 {\bm S}), \quad \lambda_5 = \mathrm{tr}({\bm W}^2 {\bm S}^2),
\end{align}
and the tensor basis \( \{{\bm T}^{(n)}\}_{n=1}^{10} \) is composed of symmetric, traceless tensors:
\begin{align} \label{eq:tb_start}
  {\bm T}^{(1)} &= {\bm S}, \\
  {\bm T}^{(2)} &= {\bm S}{\bm W} - {\bm W}{\bm S}, \\
  {\bm T}^{(3)} &= {\bm S}^2 - \frac{1}{3}{\bm I} \cdot \mathrm{tr}({\bm S}^2), \\
  {\bm T}^{(4)} &= {\bm W}^2 - \frac{1}{3}{\bm I} \cdot \mathrm{tr}({\bm W}^2), \\
  {\bm T}^{(5)} &= {\bm W}{\bm S}^2 - {\bm S}^2{\bm W}, \\
  {\bm T}^{(6)} &= {\bm W}^2{\bm S} + {\bm S}{\bm W}^2 - \frac{2}{3}{\bm I} \cdot \mathrm{tr}({\bm S}{\bm W}^2), \\
  {\bm T}^{(7)} &= {\bm W}{\bm S}{\bm W}^2 - {\bm W}^2{\bm S}{\bm W}, \\
  {\bm T}^{(8)} &= {\bm S}{\bm W}{\bm S}^2 - {\bm S}^2{\bm W}{\bm S}, \\
  {\bm T}^{(9)} &= {\bm W}^2{\bm S}^2 + {\bm S}^2{\bm W}^2 - \frac{2}{3}{\bm I} \cdot \mathrm{tr}({\bm S}^2{\bm W}^2), \\
  {\bm T}^{(10)} &= {\bm W}{\bm S}^2{\bm W}^2 - {\bm W}^2{\bm S}^2{\bm W}. \label{eq:tb_end}
\end{align}
Applying Eqs.~(\ref{eq:nonlocal_solution},\ref{eq:tensor_expansion}) to the PH 
assumed to depend on the VGT only, we arrive at:
\begin{multline} \label{eq:TBNN_approx}
    H_{ij}({\bm x}) = \mathrm{P.V.}\!\iiint \frac{\delta_{ij} - 3\bar{r}_i \bar{r}_j}{2\pi |{\bm r}|^3} Q({\bm x} + {\bm r}) \, d{\bm r} \\
    \approx \sum_{n=1}^{10} g^{(n)}(\lambda_1, \dots, \lambda_5) T^{(n)}_{ij},
\end{multline}
where the spatial dependence of the right-hand side is retained implicitly through the spatially varying VGT, \( \lambda_k = \lambda_k({\bm A}({\bm x})) \) and \( {\bm T}^{(n)} = {\bm T}^{(n)}({\bm A}({\bm x})) \) (Eqs.~(\ref{eq:vgt_invariants}--\ref{eq:tb_end})). For simplicity of notation, and because the training data discard spatial information, we omit this dependence.

This approximation reduces the nonlocal integration in Eq.~(\ref{eq:nonlocal_solution}) to the task of approximating the (potentially complex) scalar functions \( g^{(n)} \) of locally known invariants. However, this relies on the \textit{strong} assumption that the nonlocal contributions of \( Q \) can be effectively captured through observations of the local VGT.  Supporting this expansion, numerical studies of the nonlocal integral in Eq.~(\ref{eq:nonlocal_solution}) suggest that, for extreme values of the VGT -- regions in phase space that exert the strongest influence on the pressure Hessian terms -- there is a shielding effect akin to that observed in electromagnetism. This shielding significantly reduces the effective integration scale in Eq.~(\ref{eq:nonlocal_solution}), as demonstrated in \cite{vlaykov2019small}.

Due to the spatial locality of the approximation in Eq.~(\ref{eq:TBNN_approx}), this formulation is particularly attractive for closure modeling of the VGT. Early models based on this approach analyzed DNS data to extract constant values of \( g^{(n)} \), further assuming a truncated tensor basis expansion \cite{lawson2015velocity}. This method is conceptually similar to the enrichment of the isotropic upstream condition used in the RDGF model, albeit with different underlying datasets -- DNS for \cite{lawson2015velocity} and a Gaussian field for \cite{johnson2016closure}.  

The Tensor Basis Neural Network (TBNN) was introduced by Ling, Kurzawski and Templeton for Reynolds-averaged turbulence modeling \cite{ling2016reynolds}; Tian et al.\ \cite{tian2021physics} first applied it to the Lagrangian VGT, extending the VGT modeling approaches by utilizing the full tensor basis and allowing the \( g^{(n)} \) functions to vary nontrivially as functions of the invariants. These functions were approximated in \cite{tian2021physics} using Neural Networks (NN); illustrated as a submodel in Fig.~(\ref{fig:aug_tbnn}). This formulation led to the following minimization problem over $\theta$-parameterized NN
\begin{gather} \label{eq:TBNN_minimization}
    \min_\theta \sum_{s=1}^{S} \|{\bm H}_s - \hat{\bm H}_s\|_2^2, \\
    \hat{\bm H}_s = \sum_{n=1}^{10} g_\theta^{(n)}(\lambda_1, \dots, \lambda_5){\bm T}^{(n)}. \nonumber
\end{gather}
Here, $s$ runs over $S$ samples, and \( {\bm H}_s, \hat{\bm H}_s \) are the ground truth and predicted pressure Hessian for the $s$-th sample, respectively.  An important feature of this formulation is that, regardless of the specific model used for \( g^{(n)}_\theta \), the tensor basis structure ensures that the predicted tensor \( \hat{\bm H} \) is symmetric, traceless, and rotationally equivariant.

The results of \cite{tian2021physics} demonstrated that the expressive NN representation of \( g^{(n)} \) trained effectively and delivered accurate predictions. State-of-the-art performance was achieved across various relevant physical metrics, including eigenvector alignments, \( Q\text{-}R \) conditional mean tangents (CMTs), and \textit{a posteriori} tests, where an initially Gaussian fields evolve into fully turbulent ones with markedly non-Gaussian statistics. The characteristic teardrop-shaped \( Q\text{-}R \) probability distributions (PDFs) served as evidence of the model's ability to capture key turbulence dynamics.  

Following the marked success of the TBNN, subsequent work extended the formulation to enable generalization across Reynolds numbers \cite{buaria2023forecasting}. Additionally, \cite{parashar2020modeling} highlighted the sensitivity of the TBNN architecture to the choice of normalization. For structural and physical consistency, we adopt the physics-inspired normalization scheme proposed in \cite{tian2021physics}, which is detailed below.

\section{Methodology of \modelabbr} \label{sec:methods}
To motivate the Lagrangian Attention Tensor Network (\modelabbr), consider the information lost in the Markovian approximation central to the TBNN, Eq.~(\ref{eq:TBNN_approx}), which replaces a spatially nonlocal integral by a function of the current, local VGT. This reduction induces degeneracy: similar values of the VGT can correspond to very different PH terms. Subsection~\ref{sec:memory_study} shows that the Lagrangian history of the VGT resolves a large part of this ambiguity.

Motivated by previous phenomenological models utilizing the deformation tensor, we condition the closure on the recent Lagrangian history of the VGT as well as on its current value:
\[
\langle \mathbf{H}(t) | \mathbf{A}(t) \rangle \to \langle \mathbf{H}(t) | \mathbf{A}(t), \mathbf{A}(t-\Delta), \dots, \mathbf{A}(t-M\Delta) \rangle.
\]

In the remainder of this section, we first describe, in Subsection \ref{sec:time-delay}, how the history is encoded into a small set of memory features; in Subsection \ref{sec:memory_study}, the single-time experiments that fix the history length and the structure of the kernels; in Subsection \ref{sec:injection}, how the features enter the tensor basis network; in Subsection \ref{sec:invariance}, the symmetry properties of the result; and in Subsection \ref{sec:training}, the data and the two-stage training.

\subsection{Memory Features from a Time-Delay Contraction}\label{sec:time-delay}

Motivated by the deformation tensor considerations, i.e., the formal solution to Eq.~(\ref{eq:deformation_tensor_ode}),
\begin{align}\label{eq:deformation_tensor_sol}
  {\bm D}\left(t; {\bm A}(t \leftarrow t-\tau)\right) = \text{Texp}\!\left(\int\limits_{t-\tau}^t dt' {\bm A}(t')\right) \nonumber \\
  = \lim_{M\to\infty}\prod_{m=0}^M \exp\left(\Delta {\bm A}(t_m)\right),
\end{align}
where $\Delta=\tau/M,\ t_m=t-m\Delta$, and $\text{Texp}$ denotes the time-ordered exponential, we introduce a parameterizable approximation. Each factor on the right-hand side of
Eq.~(\ref{eq:deformation_tensor_sol}) can be expanded as
\begin{multline*}
  e^{{\bm A}(t_m) \Delta} = \sum_{p=0}^\infty \frac{\left( {\bm A}(t_m) \Delta \right)^p}{p!} \\
  = {\bm I} + {\bm A}(t_m) \Delta + \mathcal{O}\big((\Delta\norm{{\bm A}})^2\big),
\end{multline*}
and for $\Delta\norm{{\bm A}} \ll 1$ the leading-order approximation of the deformation tensor is
\begin{multline} \label{eq:deformation_tensor_linear_approx}
  \hat{\bm D}(t; {\bm A}(t \leftarrow t-\tau)) \\
  = {\bm I} + \Delta \sum_{m=0}^M {\bm A}(t_m) + \mathcal{O}\big((\tau\norm{{\bm A}})^2\big).
\end{multline}

Rather than inserting $\hat{\bm D}$ itself into the TBNN, we generalize Eq.~(\ref{eq:deformation_tensor_linear_approx}) to a set of $L$ learned weightings of the history, the memory features
\begin{equation} \label{eq:temp_charac_def}
    c^{(\ell)} = \sigma \left( \sum_{m=0}^{M} {\bm K}^{(m,\ell)} : {\bm A}^{(m)} \right), \qquad \ell = 1,\dots,L,
\end{equation}
where ${\bm A}^{(m)} \doteq {\bm A}(t_m)$, ${\bm K}^{(m,\ell)}$ are learnable $3\times3$ kernels, ${\bm K}:{\bm A} = K_{ij}A_{ij}$, and $\sigma$ is the logistic sigmoid. The operation is a convolution in \textit{time} only: for each feature $\ell$, every lag $m$ contributes one scalar - the full contraction of the kernel with the VGT at that lag - and the lags are summed. Note that there is no spatial filtering; the $3\times3$ index structure of ${\bm K}^{(m,\ell)}$ only weights the components of ${\bm A}^{(m)}$. Equivalently, Eq.~(\ref{eq:temp_charac_def}) is a dense mixing of the $9(M+1)$ history entries into $L$ features, constrained to act through tensor contractions.

Eq.~(\ref{eq:temp_charac_def}) differs from Eq.~(\ref{eq:deformation_tensor_linear_approx}) in two respects. First, the factor $\Delta$: with $\Delta\to0$, $M\Delta=\tau$ fixed, the sum in Eq.~(\ref{eq:temp_charac_def}) is a Riemann sum of $\int_{t-\tau}^{t}{\bm k}(t-t'):{\bm A}(t')\,dt'$ with ${\bm K}^{(m,\ell)} = \Delta\,{\bm k}^{(\ell)}(m\Delta)$, so $\Delta$ is absorbed into the learned kernels without loss of generality. Second, the identity: ${\bm K}:({\bm I}+\Delta{\bm A}) = \tr{\bm K} + \Delta\,{\bm K}:{\bm A}$, and because ${\bm A}$ is traceless the isotropic part of ${\bm K}$ never contributes to ${\bm K}:{\bm A}$; retaining ${\bm I}$ would therefore only add one free bias per feature inside $\sigma$. We omit it: the features enter the network centered on their DNS mean (Subsection \ref{sec:injection}), and with the kernel scales used here they sit in the near-linear range of $\sigma$ (standard deviation 0.21 about $\sigma(0)=1/2$, about 4\% of values saturated), where such a bias has no effect beyond a constant shift. The sigmoid bounds the features for arbitrary histories, which keeps the downstream network stable.

The length and sampling of the history, and the symmetry constraints on the kernels, are set in Subsection~\ref{sec:memory_study}.

\subsection{Memory Hyperparameters}\label{sec:memory_study}

We allowed the data to drive the remaining inductive biases of the memory. We first trained the most direct implementation on single-time (tangent) data, with unconstrained kernels and a range of history lengths and lag spacings. The memory features are concatenated with the invariants as inputs of the network for the $g$-functions, and both are optimized together on the squared error of the closure,
\begin{equation} \label{eq:tc_pred}
  \hat{\bm H} = \sum_{n=1}^{10} g_\theta^{(n)}(\lambda_1, \dots, \lambda_5, c^{(1)}, \dots, c^{(L)}) {\bm T}^{(n)}.
\end{equation}
We refer to this network as the \textit{a-priori-trained} \modelabbr. These trainings set the length of the history and the structure of the kernels.

\begin{figure}
  \centering
  \includegraphics[width=\linewidth]{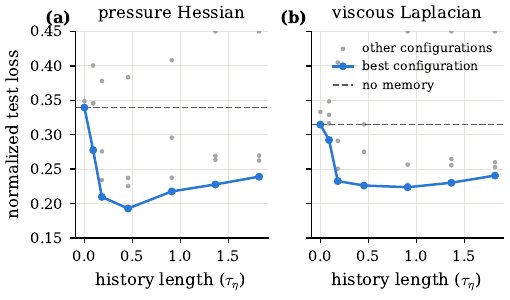}%
  \caption{Single-time test loss (mean squared error normalized by that of the zero prediction) of the a-priori-trained \modelabbr\ against the length of the history window, for the pressure Hessian (a) and viscous Laplacian (b). Zero history is the memoryless TBNN with the same network. Each grey point is one network configuration of the hyperparameter search; the line follows the best configuration at each history length.}
  \label{fig:loss_vs_history}
\end{figure}

\textit{History length.} Conditioning on the Lagrangian history lowered the single-time error substantially relative to the Markovian TBNN (Fig.~\ref{fig:loss_vs_history}), by 43\% for the pressure Hessian and by 29\% for the viscous Laplacian. The most useful history was short. For the pressure Hessian the loss was lowest at $0.45\,\taueta$ ($50\,\Delta t$) and rose for longer windows; for the viscous Laplacian it was flat between $0.45$ and $0.9\,\taueta$. We therefore use a window of $50\,\Delta t = 0.45\,\taueta$, sampled at 11 lags $5\,\Delta t$ apart, throughout. The useful memory is a fraction of a Kolmogorov time, in line with the recent-deformation picture \cite{chevillard2006lagrangian}. Longer windows lose accuracy at a fixed training budget. Appendix \ref{sec:app_state_memory} gives a model-free demonstration that memory in DNS is, in fact, longer than this window and depends on the state, but this model was unable to leverage the additional marginal information of long memory.

\begin{figure}
    \centering
    \includegraphics[width=\linewidth]{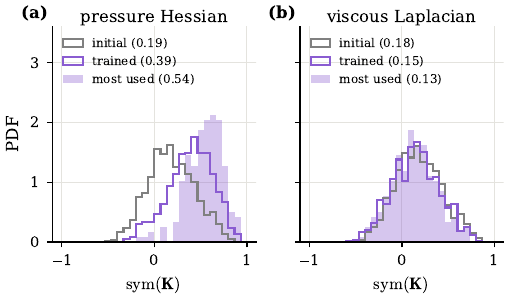}%
    \caption{Symmetry $\mathrm{sym}({\bm K})$ of the memory kernels of an a-priori-trained \modelabbr\ (unconstrained kernels, 57,600 optimizer steps), kernel by kernel: at initialization (dotted), after training (outline), and for the 20\% of kernels the network uses most (filled; \textit{usage} is the norm of the kernel times the norm of its feature's input weights).}
    \label{fig:kernel_symmetry}
\end{figure}

\textit{Kernel symmetry.} With the symmetry measure
\begin{equation}
  \mathrm{sym}(\textbf{M}) = \frac{\norm{\textbf{S}_M} - \norm{\textbf{A}_M}}{\norm{\textbf{S}_M} + \norm{\textbf{A}_M}},
\end{equation}
($\textbf{S}_M$, $\textbf{A}_M$ the symmetric and antisymmetric parts; $+1$ for a symmetric, $-1$ for an antisymmetric matrix), the unconstrained kernels of the pressure-Hessian network became markedly more symmetric during training (Fig.~\ref{fig:kernel_symmetry}): the mean moved from 0.19 at initialization to 0.39, and to 0.54 for the kernels the network uses most, and the symmetry was still strengthening at the end of training. Since ${\bm K}={\bm K}^T$ implies ${\bm K}:{\bm A} = {\bm K}:{\bm S}$, a symmetric kernel reads only the strain-rate history, so the network predicts the pressure Hessian predominantly from the history of the strain rate. The viscous kernels show no preference (0.18 to 0.15). In the \modelabbr\ we therefore build this split in: half of the $L$ kernels are constrained to be symmetric, reading the strain-rate history ${\bm S}^{(m)}$, and half antisymmetric, reading the rotation-rate history ${\bm W}^{(m)}$, ${\bm K}^{(m,\ell)} = \pm{\bm K}^{(m,\ell)T}$.

These single-time gains, however, did not carry over to the integrated model. Integrated in time, the a-priori-trained \modelabbr\ produced worse long-time statistics than the memoryless network it extends, and it was less stable (Appendix \ref{sec:app_apriori}). The network that best exploits the memory at single times is therefore not the model we want; Subsection~\ref{sec:injection} describes the construction we use instead.

\subsection{Memory Incorporation}\label{sec:injection}

\begin{figure*}
    \centering
    \includegraphics[width=\textwidth]{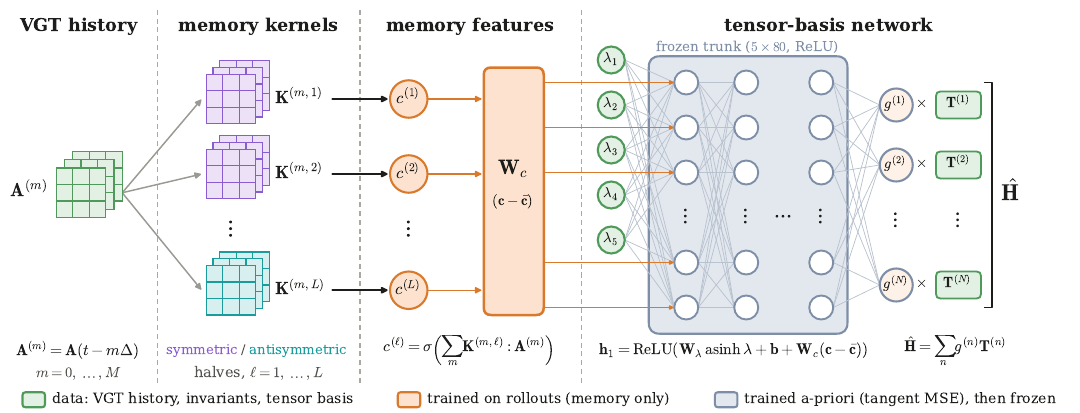}
    \caption{The \modelabbr. Left: the VGT history ${\bm A}^{(m)}={\bm A}(t-m\Delta)$ is contracted with learned kernels ${\bm K}^{(m,\ell)}$ (half symmetric, half antisymmetric) and summed over lags; a sigmoid gives the memory features $c^{(\ell)}$. Right: a tensor basis network trained on single-time data and then frozen; the centered memory features enter the pre-activation of its first hidden layer through ${\bm W}_c$, Eq.~(\ref{eq:injection}), and the network's output coefficients $g^{(n)}$ weight the tensor basis ${\bm T}^{(n)}$. Only the kernels and ${\bm W}_c$ (orange) are trained on model rollouts; the coefficients $g^{(n)}$ are produced by frozen weights but carry the memory's influence.}
    \label{fig:aug_tbnn}
\end{figure*}

The \modelabbr\ proper (Fig.~\ref{fig:aug_tbnn}) keeps a memoryless TBNN intact and lets the \textit{memory act through it}. Our \textit{trunk} -- a TBNN with $g$-functions represented by a multi-layer perceptron (5 hidden layers of 80 ReLU units, invariants compressed by $\mathrm{asinh}$, Subsection~\ref{sec:training}) -- is first trained on single-time data and then frozen. The memory features are added to the pre-activation of its first hidden layer,
\begin{equation} \label{eq:injection}
  {\bm h}_1 = \mathrm{ReLU}\!\left({\bm W}_\lambda\,\mathrm{asinh}\,\boldsymbol\lambda + {\bm b} + {\bm W}_c\left({\bm c}-\bar{\bm c}\right)\right),
\end{equation}
where ${\bm W}_\lambda$, ${\bm b}$ and all later layers are the frozen TBNN, $\bar{\bm c}$ is the mean of the features over DNS histories, and ${\bm W}_c$ (an $80\times L$ matrix) together with the kernels ${\bm K}^{(m,\ell)}$ are the only trained parameters. The memory therefore acts through the trained network, moving each sample to a different point of the TBNN's own first-layer representation; the frozen layers turn this shift into a change of all the $g$-functions. At initialization ${\bm W}_c$ is small enough that the \modelabbr\ reproduces the TBNN to within 1\% (relative $L_2$ difference of the predicted pressure Hessian on DNS histories). The same construction is used for the viscous Laplacian, with the 16-element basis of Appendix \ref{sec:VT}.

Appendix \ref{sec:app_alternatives} compares this choice with alternative points of entry for the memory.

\subsection{Symmetries}\label{sec:invariance}

Galilean invariance is satisfied trivially: the model's inputs are velocity gradients only, which are unchanged by a uniform translation or boost of the frame. Rotational equivariance is not. A tensor basis model is exactly equivariant -- rotating ${\bm A}$ rotates $\hat{\bm H}$ -- because the $g$-functions depend only on invariants. The memory features, Eq.~(\ref{eq:temp_charac_def}), contract the history with fixed kernels and are not invariant, so the \modelabbr\ breaks the exact equivariance of the TBNN: rotating the history changes the predicted closure by a median 23\% of its magnitude for the pressure Hessian and 33\% for the viscous Laplacian (Appendix~\ref{sec:equivariance}). An equivariant memory is possible, built from the history of the invariants and of mixed invariants of ${\bm A}(t)$ and ${\bm A}(t-m\Delta)$, but in our tests it gave worse integrated statistics than the non-invariant form, so we keep the latter.

\subsection{Ground Truth Data \& Training} \label{sec:training}

Our primary dataset is derived from Direct Numerical Simulation (DNS) of forced isotropic turbulence, governed by the NS equations (\ref{eq:NS}), on an Eulerian grid with $Re_\lambda \approx 240$. The numerical implementation of the Eulerian DNS is on a $1024^3$ mesh over the three-dimensional box, $\Omega=[0,2\pi]^3$. The DNS code follows a standard fully dealiased pseudospectral algorithm using a combination of truncation and phase shifting, for the incompressible Navier-Stokes equations \cite{Daniel18,Livescu00}, with Adams-Bashforth-Moulton method used for time advancement.
A large-scale linear forcing term is applied to the simulation to prevent turbulence from decaying \cite{lundgren2003linearly,petersen2010forcing}. The forcing method allows the specification of the Kolmogorov scale, $\eta$, at the onset and ensures that it remains close to the specified value, which for the dataset used is $\eta k_{max}=1.5$, where $k_{max}=\frac{\sqrt{2}}{3}N$ is the maximum resolved wavenumber for a grid size of $N^3$. Using the Eulerian velocity field, we evolve Lagrangian tracer particles initialized at random locations \cite{tian2021physics}. This provides 131,072 single-particle trajectories, each recorded for 1000 timesteps with $\Delta t = 3 \times 10^{-4}$, along which we store the VGT, Pressure Hessian (PH), and Viscous Laplacian (VL).

\textit{Time scales.} We measure time in Kolmogorov units, $\taueta = (2\langle {\bm S}:{\bm S}\rangle)^{-1/2} = 0.0330 = 110\,\Delta t$. A trajectory of 1000 steps therefore spans $9.1\,\taueta$, about 0.15 of a large-eddy turnover time. Following \cite{tian2021physics}, the VGT is normalized as $\tau{\bm A}$ with $\tau = \langle {\bm S}:{\bm S}\rangle^{-1/2} = \sqrt{2}\,\taueta$, and the invariants and tensor basis are computed from the normalized VGT.

\textit{Input conditioning.} The invariants of the VGT span many orders of magnitude: on the normalized data their medians are $O(0.1)$ while their maxima reach $O(10^2)$, and the higher-order tensor basis elements reach $O(10^3)$. Fed untransformed to the network, these heavy tails dominate the gradient; we found the TBNN trained with the published recipe of \cite{tian2021physics} to be seed-unstable on this data (in two of three seeds its pressure-Hessian test error exceeded that of the zero prediction). We therefore compress the invariants with $\lambda_k \mapsto \mathrm{asinh}(\lambda_k)$ before the first layer. Unlike a logarithm, $\mathrm{asinh}$ is defined for invariants that are negative or change sign ($\lambda_2$ to $\lambda_5$): it is odd, linear near zero and logarithmic in the tails. This single change makes training stable and reproducible, and it is what allows a nonlinear network with the full 16-element basis for the viscous Laplacian (Appendix \ref{sec:VT}) instead of the linear diffusion model of \cite{tian2021physics}. With it, the TBNN architecture of \cite{tian2021physics} reaches the same single-time error as our 5$\times$80 trunk, so the single-time improvement of our trunk over the published TBNN comes from the conditioning of the inputs rather than from the network size.

\textit{Stage 1: single-time (tangent) training.} The trunk TBNN for each of the PH and VL is trained by minimizing the squared error of the predicted closure at single times,
\begin{multline}
  \theta_{PH} = \text{arg}\min_\theta \sum_s \norm{\textbf{H}_s - \hat{\textbf{H}}_s}_2^2,\\
  \theta_{VL} = \text{arg}\min_\theta \sum_s \norm{\textbf{V}_s - \hat{\textbf{V}}_s}_2^2,
\end{multline}
on samples from the first 75\% of each trajectory (the last 25\% is the test set). The a-priori-trained \modelabbr, Eq.~(\ref{eq:tc_pred}), is trained the same way with the memory and the network optimized jointly. Hyperparameters are listed in Table~\ref{tab:hyper}.

\textit{The stochastic model.} The closures define the stochastic differential equation
\begin{equation} \label{eq:parameterized_sde}
    d\hat{\textbf{A}} = \left(\textbf{E} + \hat{\textbf{H}}_{\theta_{PH}} + \hat{\textbf{V}}_{\theta_{VL}}\right) dt + \hat{\textbf{F}}(D_s, D_a)dW,
\end{equation}
where $\hat{\textbf{F}}(D_s, D_a)dW$ is an isotropic, traceless white noise with separate strain and rotation amplitudes, as in \cite{johnson2016closure}. It represents everything the closure does not resolve, including the effect of the large-scale forcing. Its amplitudes are not tuned: they are set from the covariance of the one-step residual, ${\bm R} = {\bm A}_{\rm DNS}(t+\Delta t) - {\bm A}_{\rm model}(t+\Delta t)$ for a deterministic step from the true history, by matching $\langle {\bm R}_s:{\bm R}_s\rangle$ and $\langle {\bm R}_a:{\bm R}_a\rangle$ (symmetric and antisymmetric parts) to the noise variance per step. The comparison between the models does not depend on this calibration: scaling the noise amplitude by 0, 0.5 or 2 moves the held-out Jensen--Shannon distance (Section \ref{sec:results}) of the \modelabbr\ between 0.102 and 0.128 and that of the TBNN between 0.248 and 0.271.

\textit{Stage 2: rollout training of the memory.} With the trunk frozen, the kernels and ${\bm W}_c$ are trained autoregressively. On chaotic trajectories a trajectory-based error is unreliable -- in our tests we found it preferred sluggish dynamics -- so we use a distributional loss instead. From a window of DNS history, Eq.~(\ref{eq:parameterized_sde}) is integrated for $R=100$ steps ($0.9\,\taueta$), advancing the drift with Heun's method and adding the (additive) noise increment once per step. The ensemble of predicted states, $\{\hat{\bm A}_i\}$, is compared with the DNS states of the same particles, $\{{\bm A}_i\}$, by a sliced-Wasserstein distance \cite{bonneel2015sliced}: in one dimension the Wasserstein distance compares the sorted samples (the quantile functions) of two distributions, and the sliced distance averages it over random projections. The states are mapped to features $\boldsymbol\phi({\bm A}) = (\log\norm{{\bm A}}_F,\ {\bm b},\ q_{\bm b},\ r_{\bm b}) \in \mathbb{R}^{12}$, with ${\bm b} = {\bm A}/\norm{{\bm A}}_F$ and $q_{\bm b}, r_{\bm b}$ its invariants, which separates the intermittent magnitude from the bounded geometry. Each feature is scaled by its DNS interdecile range (${\bm D}$) and the result whitened with its DNS covariance, $\tilde{\boldsymbol\phi} = {\bm \Sigma}^{-1/2}{\bm D}^{-1}\boldsymbol\phi$, both computed from the DNS states of each batch. The loss is
\begin{multline} \label{eq:sw_loss}
  \mathcal{L} = \frac{1}{KJ}\sum_{k=1}^{K}\sum_{j=1}^{J}\Big[F^{-1}_{\hat{\bm A},{\bm u}_k}(p_j) - F^{-1}_{{\bm A},{\bm u}_k}(p_j)\Big]^2 \\
  + \lambda\,\Big\langle \mathrm{softplus}\big(\tau\norm{\hat{\bm A}_i}_F - 10\big)\Big\rangle_i,
\end{multline}
where $F^{-1}_{\cdot,{\bm u}}$ is the empirical quantile function of the projections ${\bm u}\cdot\tilde{\boldsymbol\phi}$ onto a random unit direction ${\bm u}$ ($K=256$, redrawn every step) at $J=100$ levels evenly spaced in $[0.005, 0.995]$, and the last term ($\lambda=1$) discourages states from leaving the physical range. Since $(\log\norm{{\bm A}}_F, {\bm b})$ determines ${\bm A}$, this objective targets the single-time distribution evaluated in Section~\ref{sec:results}; the evaluation tests whether a match imposed over $0.9\,\taueta$ on training windows persists to $9.1\,\taueta$ on held-out ones, and the two-time statistics are not targeted. The memory is trained with Adam (learning rate $3\times10^{-4}$, batches of 4,096 trajectories) for 36 epochs of 96 steps; we report the model at the final epoch, and give its variability over three random seeds.

\begin{table*}
\begin{ruledtabular}
\begin{tabular}{lll}
 & TBNN \cite{tian2021physics} & \modelabbr\ \\
\hline
$g$-network & $8\times40$, ReLU & $5\times80$, ReLU (frozen) \\
inputs & invariants & $\mathrm{asinh}$ invariants \\
VL closure & linear diffusion & 16-element basis \\
memory & none & $L=200$ (100 sym., 100 antisym.) \\
stage 1 & Adam $10^{-3}\!\to\!10^{-6}$, batch 4096 & Adam $2\times10^{-3}$, batch $2^{17}$ \\
stage 2 & none & memory only, rollouts of $0.9\,\taueta$ \\
noise & residual-calibrated & residual-calibrated \\
\end{tabular}
\end{ruledtabular}
\caption{\label{tab:hyper}Models and hyperparameters. History: 50 steps ($0.45\,\taueta$) sampled every $5\,\Delta t$ ($M+1=11$ lags).}
\end{table*}

\section{Numerical Experiments} \label{sec:results}
We compare the TBNN of Ref.~\cite{tian2021physics} trained with its published recipe (``TBNN'') with the \modelabbr\ of Section~\ref{sec:methods} (Fig.~\ref{fig:aug_tbnn}). For completeness, the a-priori-trained \modelabbr\ of Eq.~(\ref{eq:tc_pred}) is compared with the \modelabbr\ in Appendix \ref{sec:app_apriori}. \textit{A-priori} metrics concern single-time predictions of the closure on the test set: the normalized MSE and the alignment of the predicted and true eigenvectors. \textit{A-posteriori} metrics concern the statistics of Eq.~(\ref{eq:parameterized_sde}) integrated from DNS initial conditions.

\textit{A-posteriori protocol.} To evaluate long-term behavior of the closure models, we compare the single-time statistics of evolved trajectories to those of DNS. To obtain the modelled trajectories, we took four non-overlapping windows from the held-out final quarter of the DNS trajectories of 32,768 particles, resulting in 131,072 initial windows (the VGT and its preceding $50\,\Delta t$). From each, the SDE is integrated for 1000 steps ($9.1\,\taueta$) with its calibrated noise, using the Euler--Maruyama scheme, which in our tests gave statistics indistinguishable from those of the Heun scheme used in training (Section~\ref{sec:training}). We then collect statistics from these trajectories and compare to the statistically stationary DNS.

Topology is compared in the normalized $q$-$r$ plane \cite{das2019reynolds,das2020characterization},
\begin{equation}
    q = -\frac{1}{2}\tr\left[{\bm b}^2\right], \quad
    r = -\frac{1}{3}\tr\left[{\bm b}^3\right], \quad {\bm b} = \frac{{\bm A}}{\norm{{\bm A}}_F},
\end{equation}
which is compact (useful under heavy-tailed statistics) and separates topology from magnitude. As a scalar summary of the distribution similarity, we use the Jensen--Shannon distance \cite{lin1991divergence} between the $q$-$r$ histograms of model and DNS (Appendix \ref{sec:app_metric}). Its value between two random halves of the DNS ensemble, 0.076, is the sampling floor.
Magnitude prediction is compared via longitudinal and transverse VGT component distributions, as well as $\norm{{\bm A}}_F$.
Trajectories that leave the physical range, $\tau\norm{{\bm A}}_F > 20$ (the largest value in the DNS is 15.2), are counted as not surviving: at $9.1\,\taueta$ these are 0.7\% of the TBNN's trajectories and 0.03\% of the \modelabbr's. Rollout-trained models are scored at their final training epoch. Performance variability over three random seeds for each model is given in Appendix \ref{sec:app_seeds}.

\subsection{\textit{A-Priori} Results}\label{sec:apriori}

\begin{figure}
  \centering
  \includegraphics[width=\linewidth]{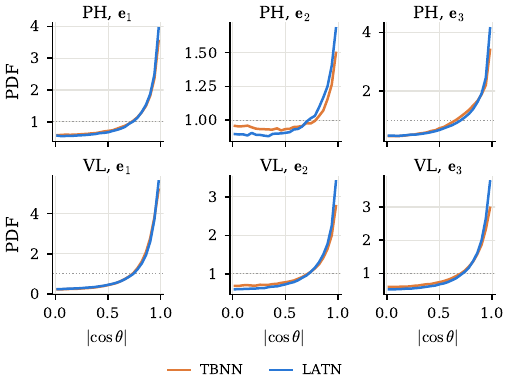}
  \caption{\textit{A-priori} alignment between the eigenvectors of the predicted and the DNS closure, $|\cos\theta|$ for ${\bm e}_1$ (most extensive), ${\bm e}_2$ and ${\bm e}_3$, for the pressure Hessian (top) and the symmetric part of the viscous Laplacian (bottom), over 655,360 test samples. Perfect prediction puts all mass at $|\cos\theta| = 1$; the dotted line is the distribution for no alignment.}
  \label{fig:priori_eig_align}
\end{figure}

\begin{table}
\begin{ruledtabular}
\begin{tabular}{lcccc}
 & \multicolumn{2}{c}{\textbf{pressure Hessian}} & \multicolumn{2}{c}{\textbf{viscous Laplacian}} \\
 & all & w/o 10 largest & all & w/o 10 largest \\
\hline
TBNN & 34.5 & 0.72 & 1.16 & 0.69 \\
trunk & 0.67 & 0.67 & 0.62 & 0.62 \\
\modelabbr & 0.71 & 0.70 & 0.95 & 0.69 \\
a-priori-\modelabbr & 0.38 & 0.36 & 0.44 & 0.43 \\
\end{tabular}
\end{ruledtabular}
\caption{\label{tab:apriori_mse}\textit{A-priori} normalized MSE (the MSE relative to that of the zero prediction) on the 655,360 test samples, over all samples and without the 10 largest per-sample errors.}
\end{table}

Eigenvector alignment is plotted as PDFs of $|\cos\theta|$, for which uncorrelated directions give a uniform density (Fig.~\ref{fig:priori_eig_align}). The \modelabbr\ aligns the eigenvectors of both closure terms marginally better than the TBNN. This alignment is a property of the frozen trunk -- not the memory kernels -- but the rollout training of the memory does preserve it. The normalized MSE (Table~\ref{tab:apriori_mse}) exhibits the same behavior: rollout training of the memory neither gains nor loses single-time accuracy, and the \modelabbr\ matches TBNN performance at single times. The TBNN's MSE, and the \modelabbr's for the viscous Laplacian, are set by a handful of extreme states; without them the networks are comparable. The a-priori-trained \modelabbr, which exploits the memory at single times, significantly outperforms both (Appendix \ref{sec:app_apriori}).

\subsection{\textit{A-Posteriori} Results}\label{sec:aposteriori}
\begin{table}
\begin{ruledtabular}
\begin{tabular}{lcccc}
 & TBNN & trunk & \modelabbr & DNS \\
\hline
JS($q$-$r$) (floor 0.076) & 0.254 & 0.128 & 0.088 & -- \\
$A_{11}$ flatness & 8.65 & 5.32 & 7.88 & 7.31 \\
$A_{11}$ skewness & $-0.37$ & $-0.42$ & $-0.43$ & $-0.49$ \\
$\langle\omega^2\rangle/2\langle{\bm S}:{\bm S}\rangle$ & 0.69 & 1.15 & 0.99 & 0.99 \\
$-\langle\tr{\bm S}^3\rangle/\langle\boldsymbol\omega\cdot{\bm S}\cdot\boldsymbol\omega\rangle$ & 0.73 & 0.44 & 0.55 & 0.76 \\
$T_I$ strain & 4.71 & 3.83 & 3.84 & 1.97 \\
$T_I$ rotation & 4.97 & 5.03 & 4.70 & 4.76 \\
\end{tabular}
\end{ruledtabular}
\caption{\label{tab:aposteriori}Quantitative summary of the \textit{a-posteriori} results across models. The Jensen--Shannon distance, the $A_{11}$ moments, and the ratios of enstrophy to strain and of strain self-amplification to vortex stretching (1 and $3/4$ in homogeneous turbulence \cite{betchov1956inequality}) are single-time statistics after $9.1\,\taueta$. The integral times $T_I$ of the strain and rotation rates, the time integrals of their Lagrangian autocorrelations (Fig.~\ref{fig:autocorr}) up to the first zero crossing, are measured along model trajectories between $4.5$ and $18\,\taueta$ and given in $\taueta$. ``Trunk'' is the frozen TBNN inside the \modelabbr, integrated without memory.}
\end{table}

\begin{figure*}
  \centering
  \includegraphics[width=\textwidth]{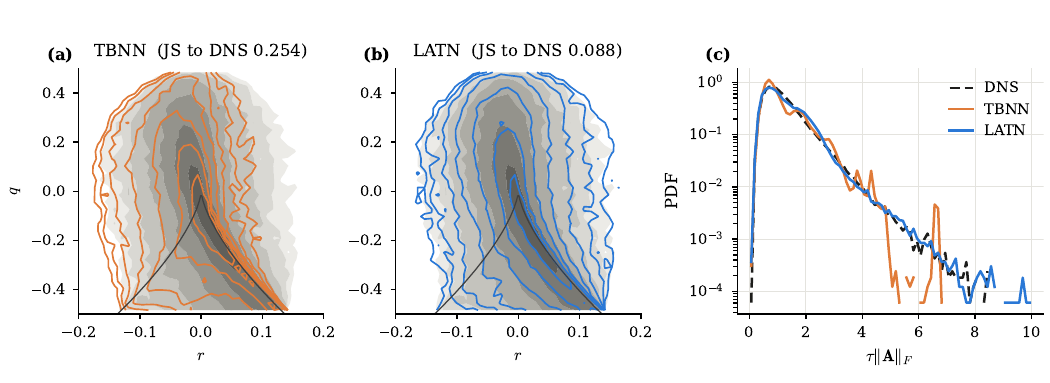}
  \caption{Normalized $q$-$r$ distributions after $9.1\,\taueta$ for the TBNN (a) and the \modelabbr\ (b), contours on the same density levels as the DNS (grey bands; the curve is the zero-discriminant line), and the distribution of the magnitude $\tau\norm{{\bm A}}_F$ (c). 131,072 initial conditions; Jensen--Shannon distances to the DNS in the panel titles.}
  \label{fig:qr_pdf}
\end{figure*}

The \modelabbr\ substantially improves the integrated statistics over the TBNN. Its $q$-$r$ distribution is within 0.012 of the sampling floor (Jensen--Shannon distance 0.088, full ensemble), against 0.254 for the TBNN (Fig.~\ref{fig:qr_pdf}). On the held-out half of the initial conditions, each of its three training seeds (0.102--0.108) is closer to the DNS than every seed of the TBNN (0.139--0.373; Appendix~\ref{sec:app_seeds}). The TBNN distorts the teardrop; the \modelabbr\ reproduces its shape, and its magnitude distribution follows the DNS to $\tau\norm{{\bm A}}_F \approx 9$, where the TBNN's shows a spurious peak near 6.5 and no states beyond 7.

Table~\ref{tab:aposteriori} separates the two ingredients of the improvement over the TBNN. The trunk alone, with conditioned inputs and a full viscous basis, already halves the $q$-$r$ distance but loses intermittency (flatness 5.3) and overproduces enstrophy relative to strain. The memory restores the intermittency and the balance of enstrophy and strain and brings the $q$-$r$ distribution to the floor.

\begin{figure}
  \centering
  \includegraphics[width=\linewidth]{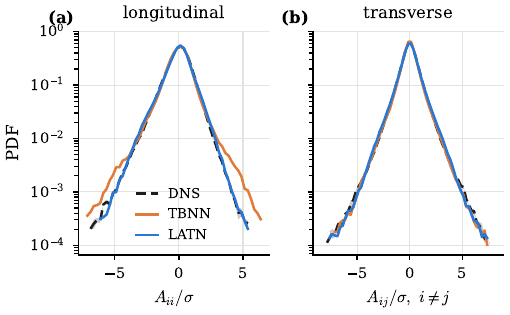}
  \caption{PDFs of the longitudinal ($A_{11}, A_{22}, A_{33}$, pooled) and transverse (off-diagonal, pooled) VGT components after $9.1\,\taueta$, normalized by the DNS standard deviation. 131,072 initial conditions, i.e.\ 393,216 and 786,432 samples; bins with fewer than 20 samples are omitted, and the grey band is the spread between two random halves of the DNS ensemble.}
  \label{fig:vgt_component_pdfs}
\end{figure}

\begin{figure}
  \centering
  \includegraphics[width=\linewidth]{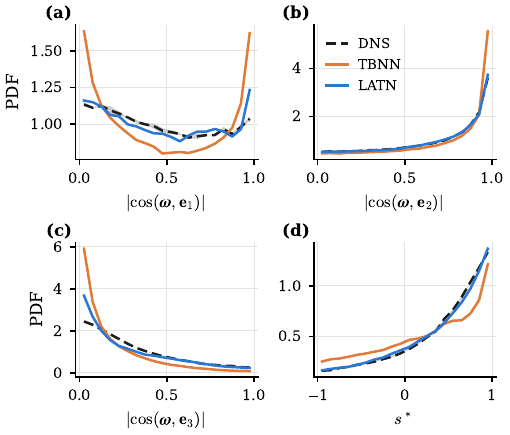}
  \caption{Alignment of the vorticity with the strain-rate eigenvectors (a--c) and the strain-state parameter $s^* = -3\sqrt{6}\,\lambda_1\lambda_2\lambda_3/|{\bm S}|^3$ \cite{lund1994improved} (d) after $9.1\,\taueta$. Grey band: DNS split-half spread.}
  \label{fig:vort_align}
\end{figure}

The component PDFs (Fig.~\ref{fig:vgt_component_pdfs}) show the intermittency directly. The \modelabbr\ follows the DNS longitudinal tails within the sampling spread down to the omitted bins. The TBNN's longitudinal tails are too heavy on both sides beyond three standard deviations. Both models reproduce the transverse PDFs. The classical alignment statistics (Fig.~\ref{fig:vort_align}) are reproduced by the \modelabbr\ more closely than by the TBNN: preferential alignment of the vorticity with the intermediate eigenvector ${\bm e}_2$, and the DNS distribution of the strain state $s^*$, which the TBNN distorts. The TBNN spuriously aligns vorticity with the most extensive eigenvector ${\bm e}_1$ (density 1.7 at $|\cos|\to1$ against 1.05 in the DNS); the \modelabbr\ largely removes this (1.3). Both models make the vorticity too often perpendicular to the most compressive eigenvector ${\bm e}_3$ (density at $|\cos| = 0$: TBNN 6.3, \modelabbr\ 3.8, DNS 2.5). We hypothesize that these peaks at the ends of the alignment PDFs are a structural tendency of tensor-basis closures: every tensor-basis model we have trained produces them, and the memory reduces but does not remove them.

\begin{figure*}
  \centering
  \includegraphics[width=\textwidth]{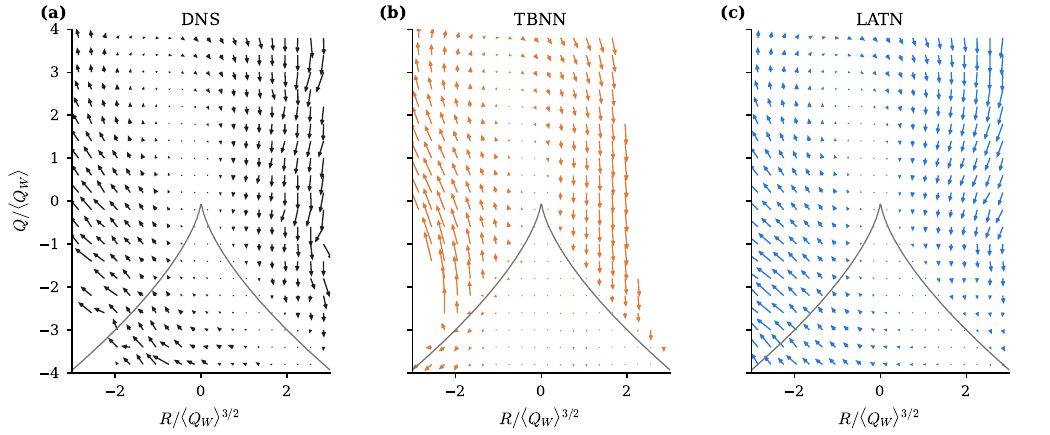}
  \caption{Conditional mean tangents $\langle (dR/dt, dQ/dt) \,|\, R, Q\rangle$ in the $Q$-$R$ plane, normalized by $\langle Q_W\rangle$, from central differences along integrated model trajectories (with the calibrated noise; steps 500--2000, i.e.\ $4.5$--$18\,\taueta$, from 16,384 DNS initial conditions) and along 16,384 DNS trajectories. Same arrow scale in every panel.}
  \label{fig:cmt}
\end{figure*}

The conditional mean tangents of the integrated models (Fig.~\ref{fig:cmt}), computed along the models' own trajectories rather than from single-time predictions on DNS states, show the dynamics behind the TBNN's topology error: its mean flow in the $Q$-$R$ plane is nearly vertical and too fast over much of the plane, while that of the \modelabbr\ resembles the DNS spiral.

\begin{figure}
  \centering
  \includegraphics[width=\linewidth]{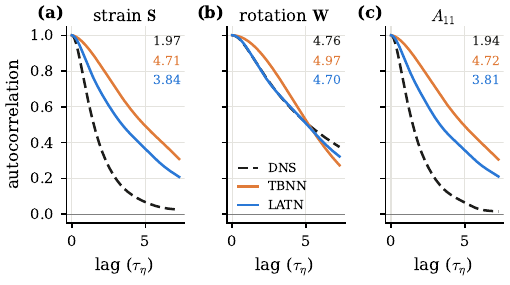}
  \caption{Lagrangian autocorrelation of the strain rate, rotation rate and $A_{11}$ along integrated model trajectories ($4.5$--$18\,\taueta$, as in Fig.~\ref{fig:cmt}) and along DNS trajectories; numbers: integral times in $\taueta$.}
  \label{fig:autocorr}
\end{figure}

The largest remaining discrepancy is in the two-time statistics (Fig.~\ref{fig:autocorr}). The rotation rate decorrelates correctly in the \modelabbr\ (integral time $4.70\,\taueta$ against 4.76 in the DNS). The strain rate, however, decorrelates about twice too slowly in every model ($3.84\,\taueta$ for the \modelabbr, 4.71 for the TBNN, 1.97 in the DNS). Table~\ref{tab:aposteriori} shows that the shortening relative to the TBNN comes from the trunk, not from the memory. A rollout objective that compares single-time distributions after $0.9\,\taueta$ does not constrain decorrelation on a $2\,\taueta$ scale.

All \textit{a-posteriori} statements in this work refer to an integration horizon of $9.1\,\taueta$, the length of the DNS trajectories. Neither model remains statistically stationary much beyond it (Section~\ref{sec:conclusion}).

\section{Conclusion} \label{sec:conclusion}
In this work, we follow the long line of blending physical insight with data analysis \cite{davidson2011voyage} to examine what the Lagrangian history of the velocity gradient contributes to closure models of its own evolution. Building on the connection between VGT evolution and fluid deformation \cite{chertkov1999lagrangian,chevillard2006lagrangian,johnson2016closure}, we augmented a tensor basis neural network with learned Lagrangian memory kernels, the Lagrangian Attention Tensor Network (LATN), and applied it to Lagrangian DNS data of forced isotropic turbulence.

Memory is strongly predictive of the closure at a single time. A history of half a Kolmogorov time lowers the error of the pressure Hessian by 43\% and markedly improves the simultaneous alignment of its eigenvectors. Kernels trained on single-time data show that the pressure Hessian is predicted predominantly from the history of the strain rate, and a model-free measurement shows that the closure's own memory shortens with the local intensity of the velocity gradient and extends beyond the history window the network can exploit (Appendix~\ref{sec:app_state_memory}).
Single-time accuracy is, however, a poor guide to the behavior of the integrated model. The network trained to exploit memory at single times consistently produces worse long-time statistics than a memoryless one (Appendix~\ref{sec:app_seeds}).
The integrated statistics improve through two ingredients. Compressing the heavy-tailed invariants with $\mathrm{asinh}$ makes single-time training of the tensor basis network stable, and this conditioned network alone halves the distance of the $q$-$r$ distribution from the DNS (Table~\ref{tab:aposteriori}). Memory added to the frozen network and trained on the model's own rollouts then brings the $q$-$r$ distribution close to the sampling floor after nine Kolmogorov times, restores an intermittency close to that of the DNS, and moves the vorticity--strain alignment toward the DNS, while the model retains the network's single-time accuracy.

Several limitations remain. The two-time statistics, computed along model trajectories that extend to $18\,\taueta$, show that the strain rate decorrelates about twice too slowly in every model. The trained memory breaks rotational equivariance: rotating the history changes the predicted closure by a substantial fraction of its magnitude (Appendix~\ref{sec:equivariance}), and the equivariant formulations we tested gave worse integrated statistics. Finally, neither model is inherently stable. Over the $9.1\,\taueta$ horizon of the DNS trajectories the memory improves stability, and in a longer integration (16,384 DNS initial conditions, not shown) the LATN's second moments stay close to the DNS up to about $10\,\taueta$, by which time the TBNN's strain is already markedly too high. Integrated further, however, the LATN fails as the TBNN does: its second moments drift away from the DNS, and about half of the trajectories of both models have diverged by $200\,\taueta$. Long-horizon stability remains an open problem for these closures and a natural target for future training objectives.

Our results additionally point to needed improvements in the learning formulation. Training the memory on single-time error buys short-time accuracy at the expense of the statistics; training it on rollouts improves the statistics without improving short-time accuracy over the TBNN. The same family of networks can reach either regime (Fig.~\ref{fig:tradeoff}). In the terms of the introduction, the LATN extends the TBNN lineage, built on short-time accuracy, toward the statistical models that reproduce Lagrangian statistics by construction \cite{carbone2024tailor,das2023data}.

The data also bear on the premise of the recent-deformation models. The history most useful to our closure spans about half a Kolmogorov time, as those models assume. However, the memory of the ground-truth (DNS) closure is longer and depends on the state. It shortens as the velocity gradient intensifies. For the pressure Hessian this mirrors, in time, the shielding of the nonlocal pressure integral, Eq.~(\ref{eq:nonlocal_solution}), whose effective range shrinks in regions of extreme velocity gradient \cite{vlaykov2019small}. Where the gradients are intense, the pressure Hessian is set by a smaller neighborhood in space and, as measured here, by a shorter stretch of its history.

The trained models and the Lagrangian DNS data used in this work are publicly available at \href{https://github.com/dlivescu/LATN}{github.com/dlivescu/LATN}.

\begin{acknowledgments}
The authors gratefully acknowledge support from: Provost startup at University of Arizona; LDRD/DR on ``MachinE Learning of Turbulence (MELT) at LANL'' (2020-21); LANL subcontract to UA ``Machine Learning of Lagrangian Turbulence'' (2022-2024). This work is co-authored by employees of Triad National Security, LLC which operates Los Alamos National Laboratory (LANL) under Contract No.89233218CNA000001 with the U.S. Department of Energy/National Nuclear Security Administration. We acknowledge supports from NNSA/LANL/ Office of Experimental Sciences and LANL's Laboratory Directed Research and development (LDRD) program, projects number 20190059DR, 20220104DR and 20240318ER, as well as computational resources from LANL's Institutional Computing (IC) program. D.L. also acknowledges support from U.S. Department of Energy, Office of Science, Office of Advanced Scientific Computing Research's Applied Mathematics Competitive Portfolios program. This document is certified by LA-UR-24-30678.
\end{acknowledgments}

\appendix

\section{Supporting Information}
\subsection{Extension to the Viscous Laplacian Term}\label{sec:VT}

In this subsection, we extend the approach described in the previous subsection (applied to the Pressure Hessian (PH) term) to the modeling of the Viscous Laplacian (VL) term. For the TBNN, this involves augmenting the tensor basis defined in Eqs.~(\ref{eq:tb_start}-\ref{eq:tb_end}) with additional skew-symmetric basis elements:
\begin{align} \label{eq:skew_tb_start}
  &\textbf{T}^{(11)} = \textbf{W}, &\textbf{T}^{(12)} &= \textbf{S}\textbf{W} + \textbf{W}\textbf{S},\\
  &\textbf{T}^{(13)} = \textbf{S}^2\textbf{W} + \textbf{W}\textbf{S}^2, &\textbf{T}^{(14)} &= \textbf{W}^2\textbf{S} - \textbf{S}\textbf{W}^2,\\
  &\textbf{T}^{(15)} = \textbf{W}^2\textbf{S}^2 - \textbf{S}^2\textbf{W}^2, &\textbf{T}^{(16)} &= \textbf{S}\textbf{W}^2\textbf{S}^2 - \textbf{S}^2\textbf{W}^2\textbf{S}. \label{eq:skew_tb_end}
\end{align}
We will build upon this extension, but first, we expand on a few subtle aspects of VL modeling.

First, in the original TBNN work \cite{tian2021physics}, it was observed that naive application of the TBNN framework to the VL term using the \( L_2 \) norm often led to overfitting, which subsequently caused stiff dynamics in the resulting stochastic differential equation (SDE). This issue was addressed by simplifying the network structure—reducing the fully nonlinear multilayer perceptron (MLP) to a linear diffusion model, a single invariant-dependent coefficient on ${\bm A}$, thereby preventing over-specialization.  

Second, a topological perspective on viscous effects \cite{das2020characterization} highlighted the dynamically nontrivial and distinct role of the VL term in the statistical evolution of the VGT topology. This insight reinforces the importance of appropriately modeling the VL term to capture its unique contributions to phase-space dynamics.

Taking these observations into account, we use a nonlinear network with the full 16-element basis for the VL; with the $\mathrm{asinh}$-compressed invariants (Section \ref{sec:training}) this trains stably. In Section \ref{sec:memory_study} we show that the learned temporal kernels for the PH and VL networks exhibit distinct properties, which supports modeling the two terms separately.

\subsection{Alternative Points of Entry for the Memory} \label{sec:app_alternatives}

We compared the injection of Eq.~(\ref{eq:injection}) with other ways of adding the memory to the frozen trunk, under the same rollout training. An additive correction to the output coefficients, $g^{(n)} = g^{(n)}_{\rm TBNN}(\boldsymbol\lambda) + \delta g^{(n)}(\boldsymbol\lambda,{\bm c})$, kept the integrated statistics stable but could not produce the intermittent tails of the VGT: the flatness of $A_{11}$ stayed near 5, against 7.5 in the DNS. Injection into a middle layer, with the later layers retrained, drifted away from the trained network and degraded the statistics. Adding nonlinear layers between ${\bm c}$ and ${\bm W}_c$ lowered the training loss but made the integrated statistics worse, while the share of the prediction that depends on the individual history fell from about 19\% to 3\%: the extra capacity was spent on the current state rather than on the history.

\subsection{Distributional Metric} \label{sec:app_metric}

The submitted version compared distributions with a relative Earth Mover's Distance against an ``ideal'' distribution; for the $q$-$r$ distributions that ideal was the DNS distribution of the same states. In the revision we use the Jensen--Shannon distance \cite{lin1991divergence}, the square root of the Jensen--Shannon divergence
\begin{gather*}
  \mathrm{JS}(P,Q) = \left[\tfrac{1}{2}\mathrm{KL}(P\,\Vert\,M) + \tfrac{1}{2}\mathrm{KL}(Q\,\Vert\,M)\right]^{1/2}, \\
  M = \tfrac{1}{2}(P+Q),
\end{gather*}
between the normalized histograms of model and DNS states on a $35\times35$ grid of the $q$-$r$ plane. It is symmetric, bounded, defined when either histogram has empty bins, and it is a metric. Its sampling floor is the value between two random halves of the DNS ensemble: 0.076 for the full ensemble of 131,072 states (0.0705 between half-ensembles). A model is indistinguishable from the DNS at this resolution when its distance reaches the floor. Model selection uses the even-indexed initial conditions and all reported numbers the odd-indexed ones.

\subsection{Variability} \label{sec:app_seeds}

\begin{table*}
\caption{\label{tab:seeds}Held-out Jensen--Shannon distance JS($q$-$r$) (floor 0.0705), $A_{11}$ flatness (DNS 7.5) and percentage of escaped trajectories after $9.1\,\taueta$ for three training seeds of each model. The seed of the reported model is marked with an asterisk.}
\begin{ruledtabular}
\begin{tabular}{lccc}
 & JS($q$-$r$) & flatness & escaped (\%) \\
\hline
TBNN & 0.373, 0.258$^*$, 0.139 & 3.4, 8.5, 34 & 0.03, 0.73, 0.64 \\
trunk & 0.140$^*$, 0.157, 0.184 & 5.2, 19, 47 & 0.02, 0.20, 0.04 \\
\modelabbr & 0.102$^*$, 0.104, 0.108 & 8.3, 8.6, 9.0 & 0.03, 0.03, 0.02 \\
a-priori \modelabbr & 0.242, 0.236, 0.276 & 103, 131, 74 & 1.0, 0.9, 0.8 \\
\end{tabular}
\end{ruledtabular}
\end{table*}

Table~\ref{tab:seeds} gives the integrated statistics of every model for three training seeds. The TBNN, the trunk and the a-priori-trained \modelabbr\ are retrained from scratch for each seed; for the \modelabbr\ the trunk is fixed and the seed sets the initialization of the memory and the sampling of its rollout training. The rollout-trained memory is reliably reproducible: its three seeds' JS($q$-$r$) lie within 0.006 of each other and below every memoryless network, with flatness of longitudinal components $A_{11}$ close to the DNS. Networks trained on single-time data vary far more. The TBNN recipe spans 0.139--0.373, and we report its median seed; its best seed matches the trunk's $q$-$r$ distance but with vast overprediction of extreme events ($A_{11}$ flatness of 34). The trunk recipe spans 0.140--0.184, and the \modelabbr\ is built on the best of its three seeds. Every seed of the a-priori-trained \modelabbr\ is further from the DNS than every seed of the trunk.

\subsection{Rotational Equivariance} \label{sec:equivariance}

\begin{figure}
    \centering
    \includegraphics[width=\linewidth]{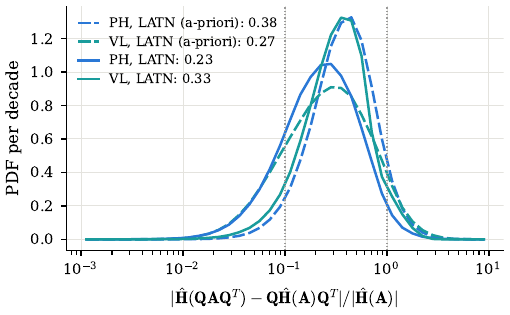}
    \caption{Violation of rotational equivariance by the \modelabbr\ (solid) and by the a-priori-trained \modelabbr\ (dashed; Appendix~\ref{sec:app_apriori}): for each of 655,360 test samples, the whole history is rotated by one random rotation ${\bm Q}$, and the relative error $|\hat{\bm H}({\bm Q}{\bm A}{\bm Q}^T) - {\bm Q}\hat{\bm H}({\bm A}){\bm Q}^T| / |\hat{\bm H}({\bm A})|$ is recorded; legend: medians. The TBNN is equivariant to floating-point precision.}
    \label{fig:equivariance}
\end{figure}

Fig.~\ref{fig:equivariance} quantifies the violation of rotational equivariance by the \modelabbr\ (Section~\ref{sec:invariance}). The error is relative to the size of the prediction: at 0.1 a rotated history changes the predicted closure by a tenth of its magnitude, a small perturbation of the same tensor; at 1 the prediction for the rotated history is no closer to the rotated prediction than the prediction is to zero. The violation shows that the memory carries frame information that the tensor basis does not see. As discussed in Sec.~\ref{sec:invariance}, we did explore equivariant formulations, but they underperformed in every one of our tests. Making the memory invariant without losing its integrated benefit is an open design problem.

\subsection{The A-Priori-Trained LATN} \label{sec:app_apriori}

\begin{figure}
  \centering
  \includegraphics[width=\linewidth]{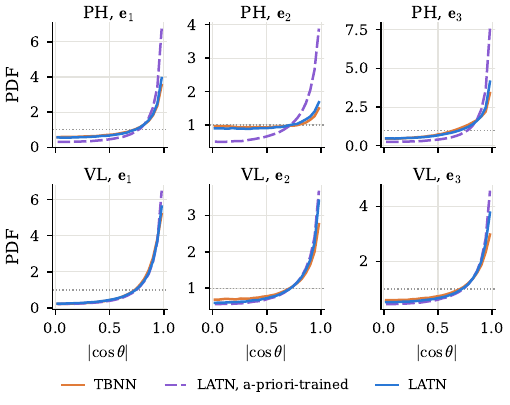}
  \caption{As Fig.~\ref{fig:priori_eig_align}, with the a-priori-trained \modelabbr\ (dashed).}
  \label{fig:app_alignment}
\end{figure}

\begin{figure}
  \centering
  \includegraphics[width=\linewidth]{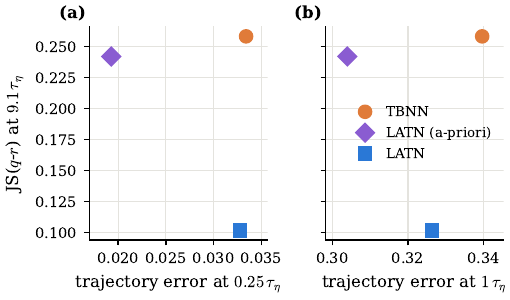}
  \caption{Trajectory error $\langle|{\bm A}-{\bm A}_{\rm DNS}|^2\rangle/\langle|{\bm A}_{\rm DNS}|^2\rangle$ of deterministic rollouts started from the exact DNS state and history (8,192 trajectories not used to train any model's single-time closure), at $0.25\,\taueta$ (a) and $1\,\taueta$ (b), against the long-time topology error (held-out Jensen--Shannon distance of the $q$-$r$ distribution after $9.1\,\taueta$). Lower left is better.}
  \label{fig:tradeoff}
\end{figure}

The a-priori-trained \modelabbr, Eq.~(\ref{eq:tc_pred}), has the trunk and the memory trained jointly on single-time data. It is the best single-time model in this work. The pressure Hessian's eigenvectors align markedly better than with the TBNN or the \modelabbr\ (Fig.~\ref{fig:app_alignment}). Its trajectories follow the DNS more closely for the first half Kolmogorov time (Fig.~\ref{fig:tradeoff}a).

Unfortunately, this accuracy degrades quickly. By $t=2\,\taueta$ it is the least accurate of the three models. Table~\ref{tab:seeds} and Fig.~\ref{fig:tradeoff} compare it with the \modelabbr\ and the TBNN at short and long times. In summary, we found training the memory on single-time error buys short-time accuracy at the cost of the long-time statistics. Its equivariance violation is comparable to the \modelabbr's (Fig.~\ref{fig:equivariance}). In earlier training, we found a single, tantalizing training of an a-priori-trained \modelabbr\ that maintained both strong short- and long-time performance, but could not reproduce the performance reliably. Reproducibility motivated us to present our more modest, \modelabbr\ formulation.

\subsection{Memory Time of the DNS Closure} \label{sec:app_state_memory}

\begin{figure}
    \centering
    \includegraphics[width=\linewidth]{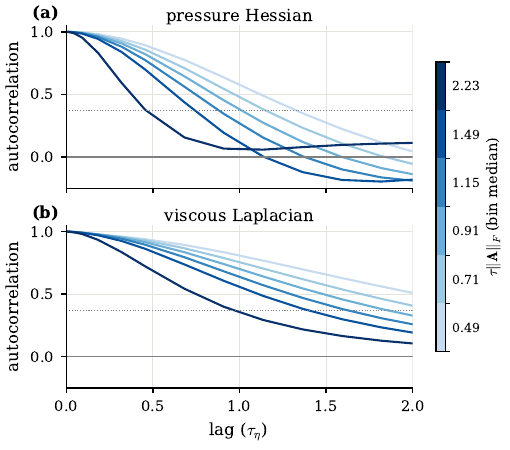}
    \caption{Memory of the true closure, measured from the DNS data alone. Correlation coefficient of the pressure Hessian (a) and viscous Laplacian (b) with their own past at lag $h$, conditioned on the current $\tau\norm{{\bm A}}_F$ (colour: six equally populated bins, labelled by their medians); dotted line, $1/e$. All 131,072 particles, sampled every $20\,\Delta t$ along their histories (5,111,808 samples).}
    \label{fig:time_kernels}
\end{figure}

Fig.~\ref{fig:time_kernels} gives a model-free measurement of how much history there is to use: the correlation coefficient between the DNS closure term at a time $t$ and at $t-h$, sampled along the history of every particle, within six equally populated bins of the current magnitude $\tau\norm{{\bm A}(t)}_F$. For the pressure Hessian the correlation falls to $1/e$ after $1.33\,\taueta$ in the calmest sixth of the states and after $0.46\,\taueta$ in the most intense, approximately as $\norm{{\bm A}}_F^{-0.68}$: the memory shortens with intensity, though more slowly than the local eddy time $\norm{{\bm A}}_F^{-1}$. The viscous Laplacian remembers about twice as long, and its memory shortens with intensity in the same way, approximately as $\norm{{\bm A}}_F^{-0.72}$. Its correlation falls to $1/e$ after $1.87\,\taueta$ in the third bin and after $0.98\,\taueta$ in the most intense, and in the two calmest bins it stays above $1/e$ beyond the longest lag measured, $2\,\taueta$. The $0.45\,\taueta$ window of the \modelabbr\ therefore spans about the whole memory of the pressure Hessian in the most intense states and about a third of it in the calm majority. Within the window the closure decorrelates little in most states, consistent with the short optimal window of Fig.~\ref{fig:loss_vs_history}.

\bibliography{refs}

\end{document}